\documentclass{icrc2009}

\usepackage{graphicx}   % for including figures
\usepackage{fixltx2e}
\usepackage{url}

\newcommand{\shorttitle}[1]%
{\markboth{Proceedings of the 31\MakeLowercase{$^{st}$} ICRC, {\L}\'{o}d\'{z} 2009}{#1} }
\begin{document}

\title{Radio Emission AIRES: Results and parameterization}
\author{\IEEEauthorblockN{Colas Rivi\`ere\IEEEauthorrefmark{1},
                          Fran\c{c}ois Montanet\IEEEauthorrefmark{1},
			  Jo\"el Chauvin\IEEEauthorrefmark{1}}
			  \\
\IEEEauthorblockA{\IEEEauthorrefmark{1}Laboratoire de Physique Subatomique et de Cosmologie, UJF, INPG, CNRS/IN2P3, Grenoble, France}
}

\shorttitle{C. Rivi\`ere, Radio Emission AIRES: Results and parameterization}
\maketitle

\begin{abstract}
Radio emission has been implemented in the simulation software AIRES by M. A. DuVernois, B. Cai and D. Kleckner to perform a full Monte Carlo simulation of the radio emission of cosmic ray extensive air shower. Different sets of showers have been simulated. We present here the dependency of the radio electric field with different parameters (primary energy, radial distance, arrival direction, etc). We stress the effect of the polarization and the arrival direction, which should have a fundamental effect on the interpretation of experimental data.
\end{abstract}

\begin{IEEEkeywords}
cosmic rays radio detection, simulation, polarization
\end{IEEEkeywords}

%%%%%%%%%%%%%%%%%%%%%%%%%%%%%%%%%%%%%%%%%%%%%%%%%%%%%%%%%%%%%%%%%%%%%%%%%%
%%%%%%%%%%%%%%%%%%%%%%%%%%%%%%%%%%%%%%%%%%%%%%%%%%%%%%%%%%%%%%%%%%%%%%%%%%
\section{Introduction}

The radio detection of cosmic rays has been first explored in the late 60's and early 70's~\cite{allan}. Thanks to progress in electronics, we currently assist to a renewal of experimental work with CODALEMA in France~\cite{CODALEMA}, LOPES in Germany~\cite{LOPES} as well as radio prototypes at the Pierre Auger Observatory, Argentina~\cite{ARENA_Benoit,ARENA_Jose}.

Various models of emission have been proposed, and several are still under development such as the Coulombian emission of a charge excess, the transverse current or the geosynchrotron emission~\cite{falcke2003dre,huege,ReAIRES,gousset,lecacheux,scholten}. If a geomagnetic effect has been observed experimentally~\cite{allan,LOPES,AP2009}, a complete model which reproduces correctly all the observations is still to be developed.
%%%%%%%%%%%%%%%%%%%%%%%%%%%%%%%%%%%%%%%%%%%%%%%%%%%%%%%%%%%%%%%%%%%%%%%%%%
%%%%%%%%%%%%%%%%%%%%%%%%%%%%%%%%%%%%%%%%%%%%%%%%%%%%%%%%%%%%%%%%%%%%%%%%%%
\section{Radio emission AIRES}
The general form of the electric field produced by an accelerated relativistic particle is given by:
\begin{eqnarray}
  \mathbf{E}&{}={}&\underbrace{\frac{e}{4\pi\epsilon_0}
    			      \left[\frac{\mathbf{n}-\mathbf{v}}
    				      {\gamma^2(1-\mathbf{v}\cdot\mathbf{n})^3R^2}\right]_{ret}}_{\textnormal{Static field}} \nonumber\\
    		      &&{+}\:\underbrace{\frac{e}{4\pi\epsilon_0c}
			      \left[\frac{\mathbf{n}\times\{(\mathbf{n}-\mathbf{v})\times\dot{\mathbf{v}}\}}
				      {(1-\mathbf{v}\cdot\mathbf{n})^3R}\right]_{ret}}_{\textnormal{Radiation field}}
  \label{jackson_formula}
\end{eqnarray}
where $\mathbf{v}$ represents the velocity of the particle, $\mathbf{n}$ is the direction and $R$ the distance from the emission point to the observer.
The synchrotron emission corresponds to an acceleration induced by a magnetic field $\dot{\mathbf{v}}=\frac{e}{\gamma m}\mathbf{v}\times\mathbf{B}$.

Radio emission has been implemented in the simulation software AIRES~\cite{AIRES} (latter called ReAIRES) by M. A. DuVernois, B. Cai and D. Kleckner~\cite{ReAIRES}. It is based on the Monte Carlo program AIRES in order to provide a realistic extensive air shower (EAS). For every electron and positron simulated, the electric field is computed based on the radiative term of Eq.~\ref{jackson_formula} using the geomagnetic field at the location of the simulation. 
The refractive index of air $n$ is approximated to unity. The static term of emission has also been implemented and has been checked to be negligible.

The AIRES thinning algorithm is used to reduce the number of particles to simulate. The minimum number of particles to simulate mainly depends on the distance of the observation point to the shower core and on the considered frequency. The thinning level has been chosen appropriately for the following simulations.

%%%%%%%%%%%%%%%%%%%%%%%%%%%%%%%%%%%%%%%%%%%%%%%%%%%%%%%%%%%%%%%%%%%%%%%%%%
%%%%%%%%%%%%%%%%%%%%%%%%%%%%%%%%%%%%%%%%%%%%%%%%%%%%%%%%%%%%%%%%%%%%%%%%%%
\section{Shower simulation}
For this analysis, showers were simulated on a virtual array of 40 antennas located up to 250~m around the shower core. The geomagnetic field is taken to Western Europe's values. The North-South (NS) axis is defined positive to the North, the East-West (EW) axis positive to the West and the vertical axis (VT) positive upwards. The shower direction is given by its direction of origin, with a zenith angle $\theta$ and an azimuth angle $\phi$ ($0^\circ\equiv\textnormal{N}$ and $90^\circ\equiv\textnormal{W}$). During the analysis, the radio signals are filtered in the 23--83~MHz band, even though the following results do not depend much on this band. We latter call \emph{signal} the amplitude of the peak of the filtered signal, which can be positive or negative.

\begin{figure*}[thbp]
  \centering
  \includegraphics[width=0.32\textwidth]{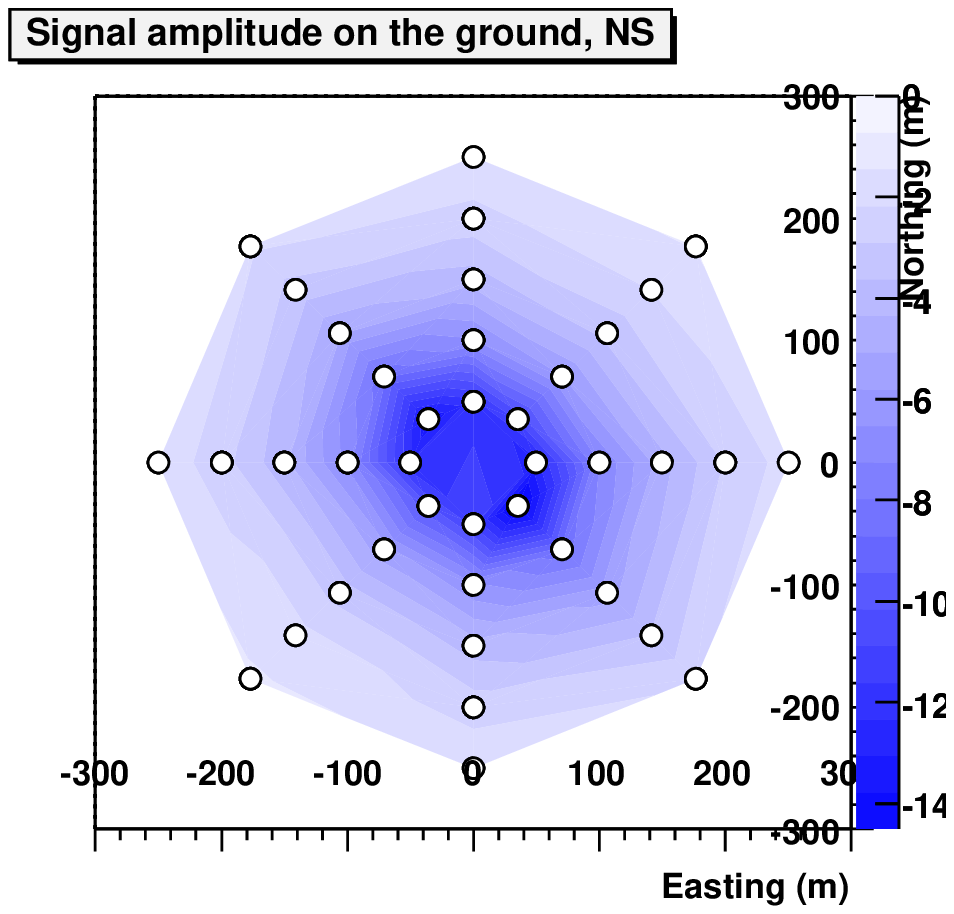}
  \includegraphics[width=0.32\textwidth]{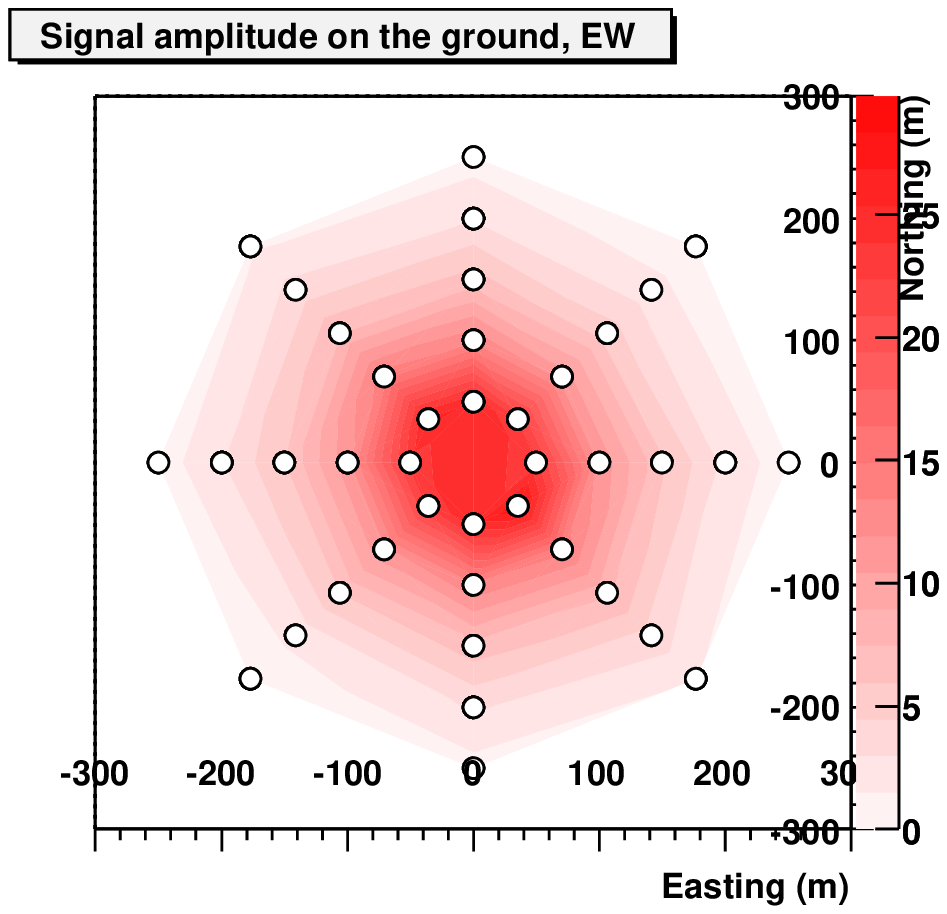}
  \includegraphics[width=0.32\textwidth]{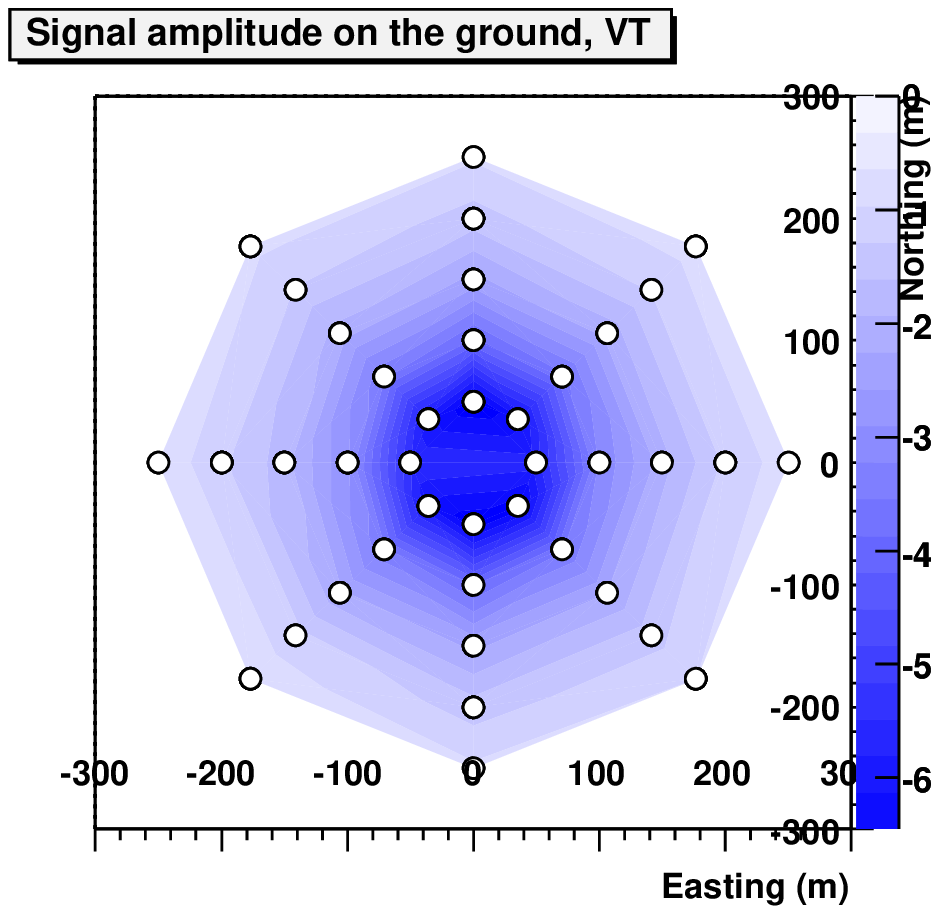}
%  \includegraphics[width=0.32\textwidth]{./img/footprint_simu_27_NS_BW}
%  \includegraphics[width=0.32\textwidth]{./img/footprint_simu_27_EW_BW}
%  \includegraphics[width=0.32\textwidth]{./img/footprint_simu_27_VT_BW}
% g b d h
  \caption{Footprint of the electric field, in each polarization, for a $10^{17}$~eV shower coming from $\theta=30^\circ$ and $\phi=45^\circ$. The dots indicate the points where the field is actually calculated. We can note a change of sign between the polarizations.}
  \label{footprints_3pol}
  \includegraphics[width=0.32\textwidth]{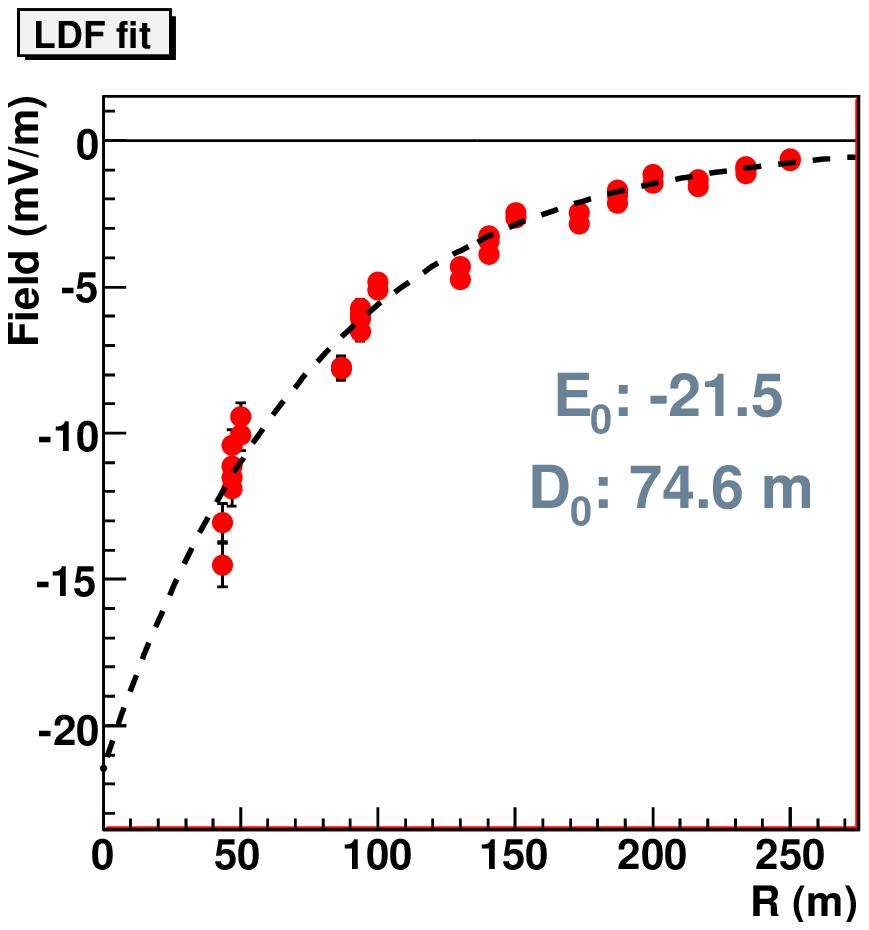}
  \includegraphics[width=0.32\textwidth]{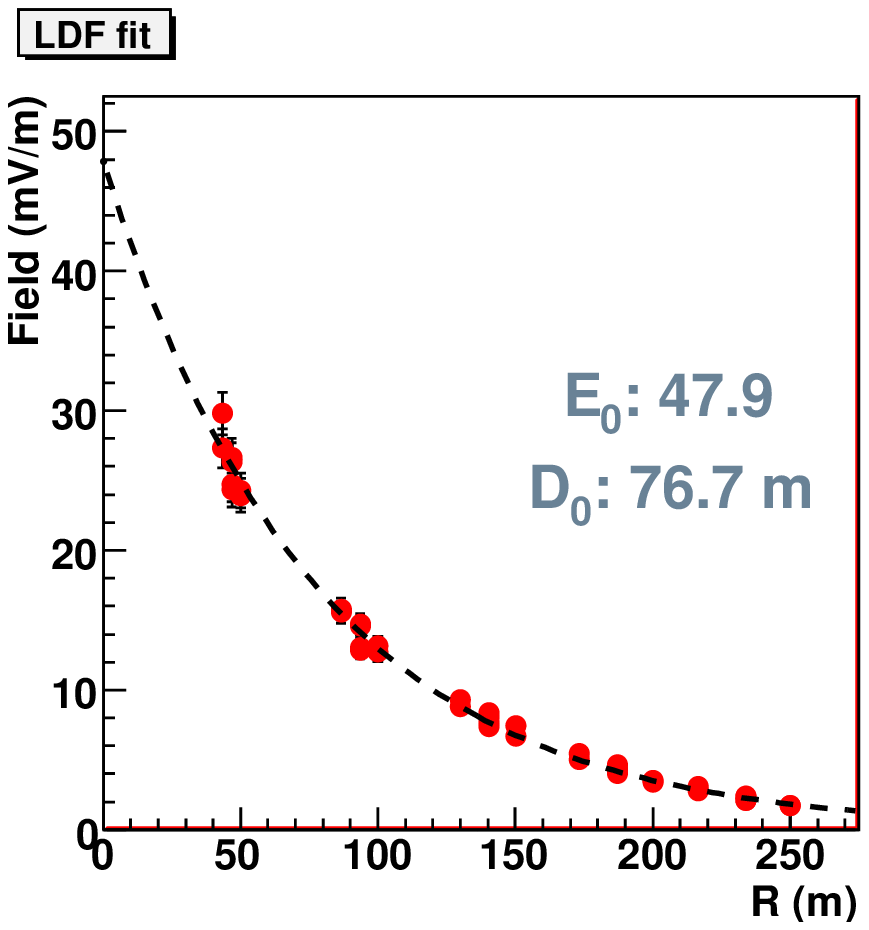}
  \includegraphics[width=0.32\textwidth]{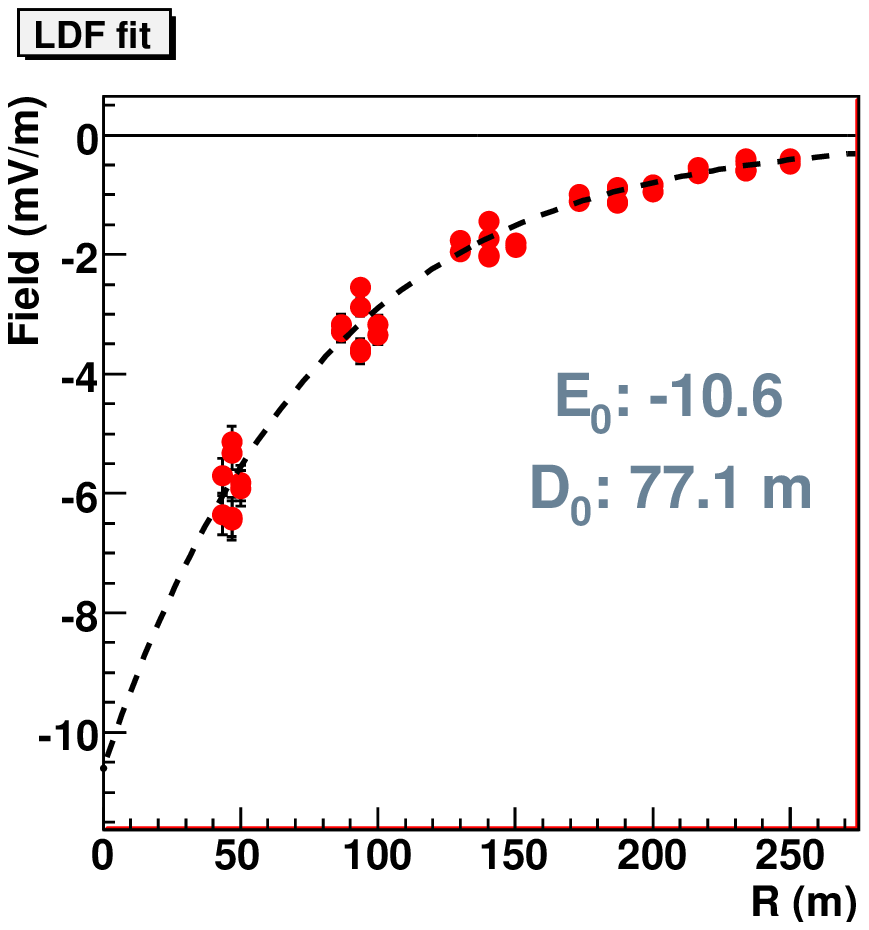}
%  \includegraphics[width=0.32\textwidth]{./img/ldf_27_NS_BW}
%  \includegraphics[width=0.32\textwidth]{./img/ldf_27_EW_BW}
%  \includegraphics[width=0.32\textwidth]{./img/ldf_27_VT_BW}
% g b d h
  \caption{Lateral distributions of the signals of Fig.~\ref{footprints_3pol}, fitted with exponential functions.}
  \label{ldf_3pol}
\end{figure*}

A typical radio footprint ($10^{17}$~eV proton induced EAS coming from $30^\circ$ zenith and $45^\circ$ azimuth, i.e. North-West) is given on Fig.~\ref{footprints_3pol}.

The electric field is generally axially symmetric around the shower axis. This is visible on Fig.~\ref{ldf_3pol}, where the signals from the same shower are plotted against the radial distance to the shower axis in the shower plane. The signals are fitted with an exponential function of the form:
\begin{equation}
  E=E_0\exp{\frac{-R}{R_0}}
  \label{exponentielle}
\end{equation}
where $E_0$ represents the field on the axis, $R$ is the radial distance and $R_0$ is the characteristic decreasing distance of the field. $E_0$ and $R_0$ are adjusted during the fit. For most of the simulated showers, the signals are well describes by the exponential function. The $R_0$ in the three polarizations are very similar.

\begin{figure*}[htbp]
  \centering
  \includegraphics[width=0.32\textwidth]{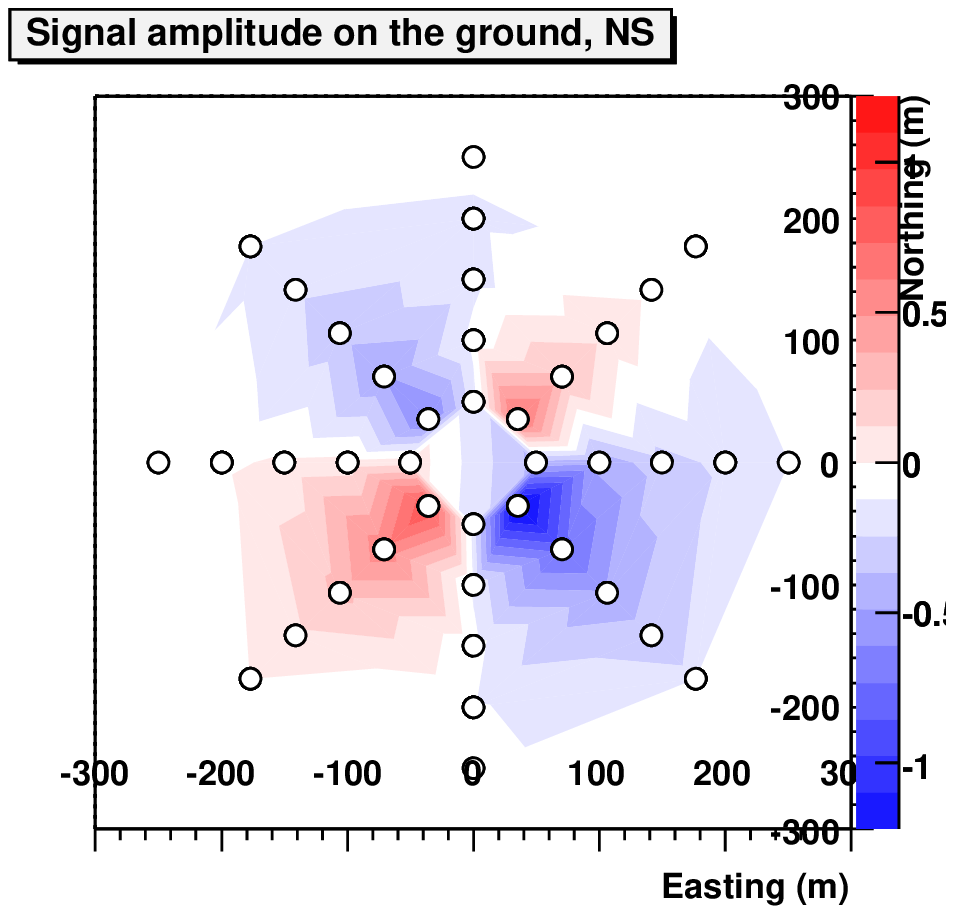}
  \includegraphics[width=0.32\textwidth]{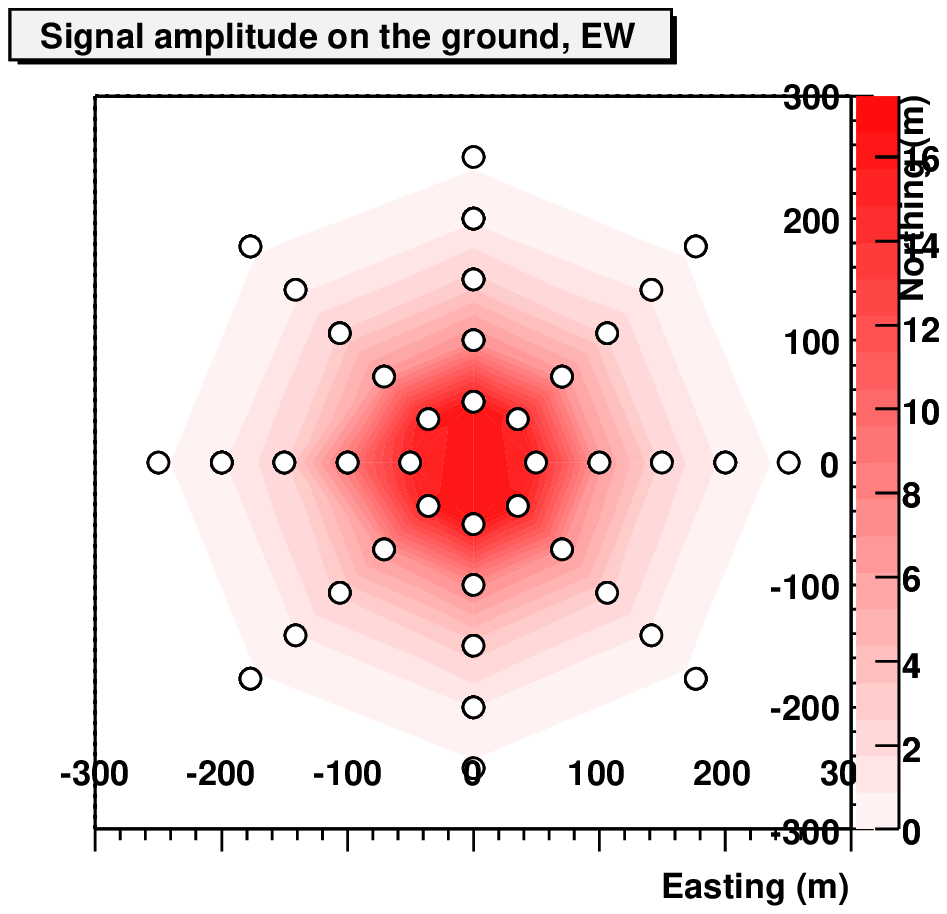}
  \includegraphics[width=0.32\textwidth]{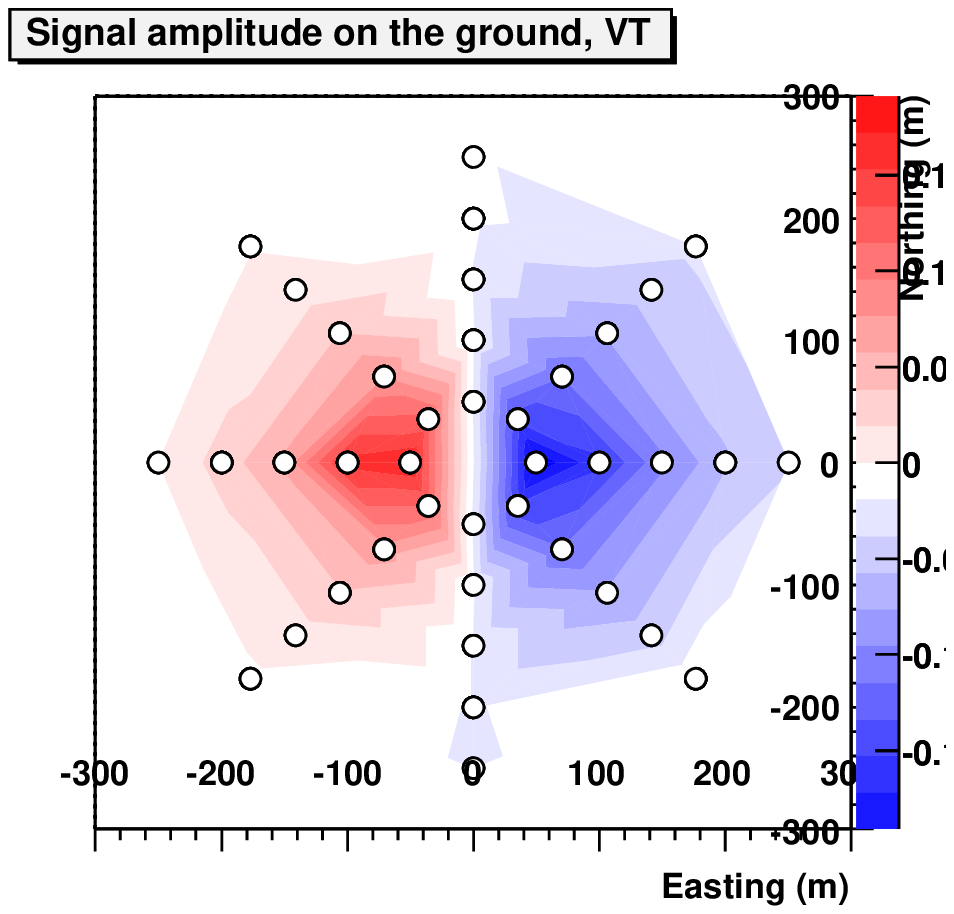}
%  \includegraphics[width=0.32\textwidth]{./img/footprint_simu_120_NS_BW}
%  \includegraphics[width=0.32\textwidth]{./img/footprint_simu_120_EW_BW}
%  \includegraphics[width=0.32\textwidth]{./img/footprint_simu_120_VT_BW}
% g b d h
  \caption{Same as Fig.~\ref{footprints_3pol}, but for a vertical shower.}
  \label{footprints_3pol_vertical}
\end{figure*}

For particular arrival directions of the shower, the footprint in a given polarization can exhibit an asymmetric pattern. A vertical shower for instance presents two changes of sign for the NS field, and one change of sign for the vertical field. However the EW field, which value is typically ten times greater than the other fields here, is axially symmetric; as illustrated on Fig.~\ref{footprints_3pol_vertical}. More generally, these kinds of patterns only appear when the signal in this polarization is very weak compared to the other polarizations, and can thus be ignored in a first order description of the electric field. It actually happens only when the projection of $\mathbf{v}\times\mathbf{B}$ in the considered polarization is very small, as discussed latter.

%%%%%%%%%%%%%%%%%%%%%%%%%%%%%%%%%%%%%%%%%%%%%%%%%%%%%%%%%%%%%%%%%%%%%%%%%%
%%%%%%%%%%%%%%%%%%%%%%%%%%%%%%%%%%%%%%%%%%%%%%%%%%%%%%%%%%%%%%%%%%%%%%%%%%
\section{Arrival direction and $\mathbf{v}\times\mathbf{B}$ polarization}
\begin{figure*}[htbp]
  \centering
  \includegraphics[width=0.32\textwidth]{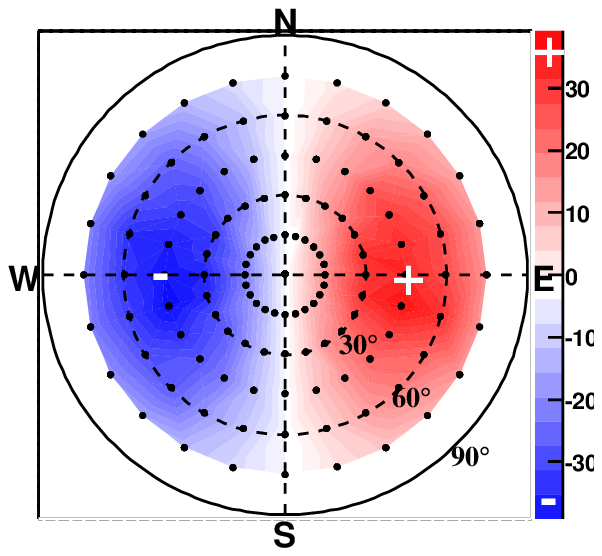}
  \includegraphics[width=0.32\textwidth]{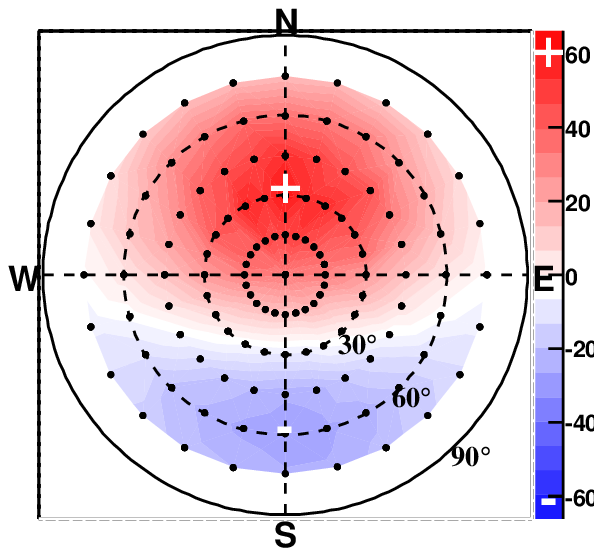}
  \includegraphics[width=0.32\textwidth]{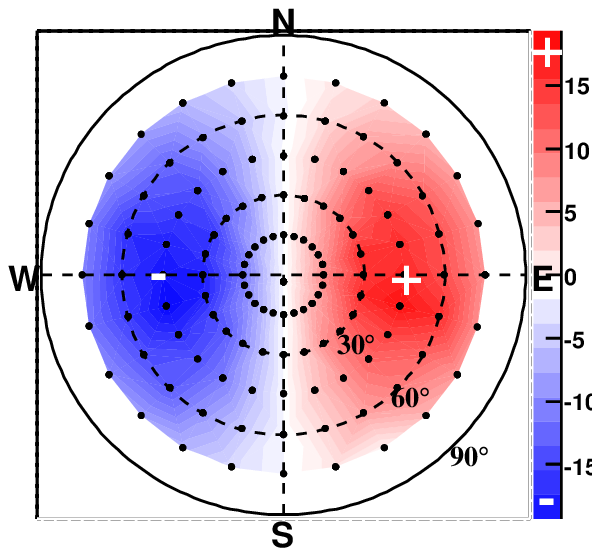}
% g b d h
  \caption{Skymaps of $E_0$ from ReAIRES: amplitude $E_0(\theta,\phi)$ as a function of the arrival direction of the simulated showers (indicated with dots), in the NS (left), EW (center) and VT (right) polarization.}
  \label{skymap_aires}
  \includegraphics[width=0.32\textwidth]{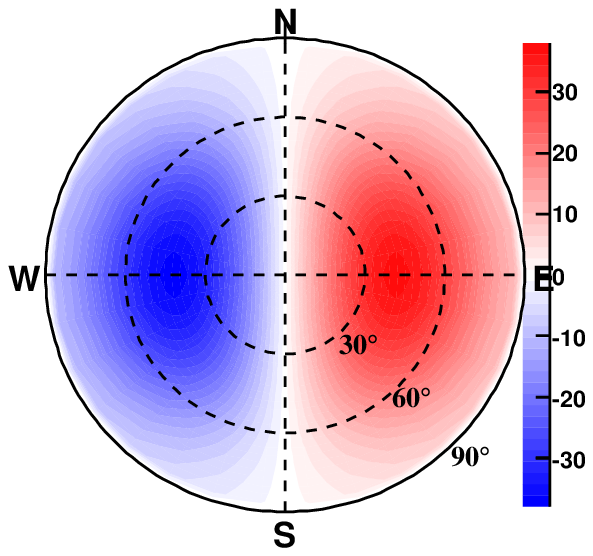}
  \includegraphics[width=0.32\textwidth]{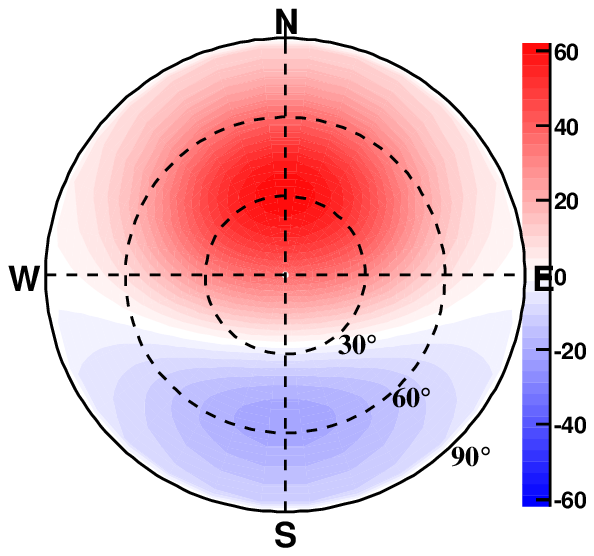}
  \includegraphics[width=0.32\textwidth]{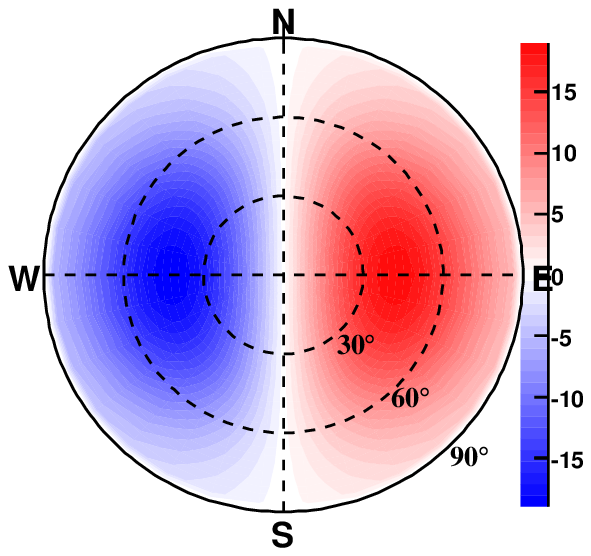}
% g b d h
  \caption{Skymaps of $F(\theta).(\mathbf{v}\times\mathbf{B})$, where $F(\theta)$ is adjusted to fit the skymap of Fig.~\ref{skymap_aires}.}
  \label{skymap_vB_F}
\end{figure*}

In order to study the effect of the arrival direction on the radio signals, sets of 121 showers of fixed energy and coming from various directions (from $\theta=0$ to $75^\circ$) have been simulated. The amplitudes $E_0$ in all three polarizations have been extracted for each shower. Their values present strong variations upon the arrival direction. This is shown for $10^{17}$~eV proton primaries as a function of the arrival direction on Fig.~\ref{skymap_aires}. The NS and VT terms are very similar\footnote{There is actually just a tangent of the geomagnetic inclination between the two.}: there are null for showers coming from North and South, are positive for showers coming from East and negative for showers coming from West. For the EW term, it is positive for showers coming from a larger North area, null on an East--West curve containing the geomagnetic field direction and negative on South.

\begin{figure}[htbp]
  \centering
  \includegraphics[width=\columnwidth,viewport=5 10 560 485,clip]{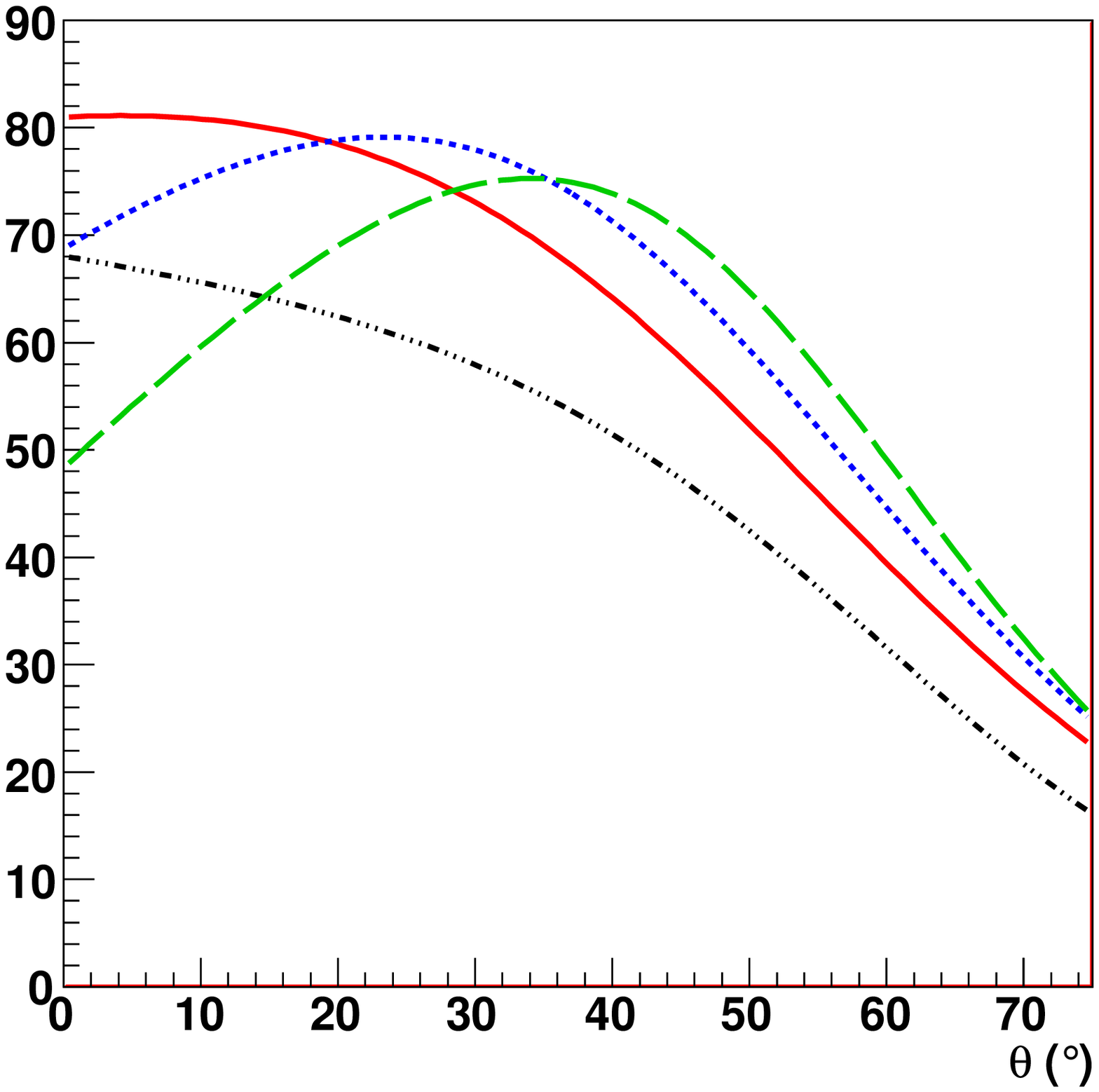}
%  \includegraphics[width=0.9\columnwidth]{./img/cFTheta_BW}
% g b d h
  \caption{$F(\theta)$ for proton primaries at different energies. The reference function is the red full line, at $10^{17}$~eV. The other functions have been rescaled by $10^2$ at $10^{15}$~eV (mixed black curve), by $10^{-2}$ at $10^{19}$~eV (dashed blue curve) and by $10^{-4}$ at $10^{21}$~eV (long dashed green curve).}
  \label{Fs_theta}
\end{figure}

This is simply explained by a linear dependence between the electric field and the vector cross product of the direction of the shower $\mathbf{v}$ and the geomagnetic field direction $\mathbf{B}$: $\mathbf{E}\propto-\mathbf{v}\times\mathbf{B}$, as predicted by some analytical models. It is the case for instance for the synchrotron radiation when the electric field is observed near the particle motion axis. Thus the skymaps presented in Fig.~\ref{skymap_aires} can be parameterized simply considering this vector cross product and a single function of the zenith angle:
\begin{equation}
  \mathbf{E}=-F(\theta).\mathbf{v}\times\mathbf{B}
  \label{parametrization_formula}
\end{equation}
The vector cross products represents the emission mechanism, and the function $F(\theta)$ contains the shower development, the longitudinal distance to the shower, the atmospheric depth, etc. This function $F(\theta)$ has been adjusted using a function of the form $A.(1+B.\theta)/(1+exp(\frac{\theta-C}{D}))$. The functions obtained are shown on Fig.~\ref{Fs_theta} for proton primaries at different energies ranging from $10^{15}$ to $10^{21}$~eV.
The $F(\theta)$ functions are almost proportional to the primary energy. The remaining differences may be interpreted in term of shower development:
\begin{itemize}
\item A low energy shower reaches its full development higher in the atmosphere, thus produces less electric field on the ground. When the zenith angle increases, the shower gets even farther, and the field is weakened.
\item A vertical high energy shower hits the ground before it reaches its full development. Increasing the zenith angle will allow the shower to develop above the ground and increase the electric field on the ground. After a given zenith angle, the shower will become too far from the observation point and the signal will start to decrease again, as for a low energy shower.
\end{itemize}
This corresponds to the tendency observed in Fig.~\ref{Fs_theta}, the global maximum tends to be reached at higher zenith angle as the energy increases.
%%%%%%%%%%%%%%%%%%%%%%%%%%%%%%%%%%%%%%%%%%%%%%%%%%%%%%%%%%%%%%%%%%%%%%%%%%
%%%%%%%%%%%%%%%%%%%%%%%%%%%%%%%%%%%%%%%%%%%%%%%%%%%%%%%%%%%%%%%%%%%%%%%%%%
\section{Overall parameterization}
\subsection{Energy dependency}
As already briefly discussed, the signal amplitude is quasi proportional to the energy of the primary cosmic ray. The simulation of a set of 41 showers in a fixed direction ($\theta=30^\circ$, $\phi=45^\circ$) with energies ranging from $10^{16}$ to $10^{20}$~eV gives the relation $\mathbf{E}\propto Energy^{1.02}$, thus we will simply consider the proportionality.

\subsection{Characteristic distance $R_0(\theta)$}
\begin{figure}[htbp]
  \centering
  \includegraphics[width=0.9\columnwidth,viewport=5 10 560 485,clip]{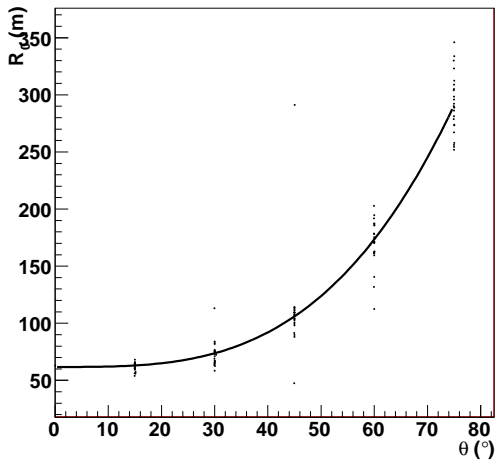}
% g b d h
  \caption{$R_0(\theta)$ the events of Fig.~\ref{skymap_aires}.}
  \label{D0_theta}
\end{figure}

As the zenith angle $\theta$ increases, the shower gets farther from the observation point and the signal lateral distribution on the ground is flattened, i.e. the characteristic distance $R_0$ of Eq.~\ref{exponentielle} increases. This effect is added to the broadening of the footprint on the ground due to the projection effect. The distribution of $R_0(\theta)$ for the $10^{17}$~eV showers used to produce the Fig.~\ref{skymap_aires} is given in Fig.~\ref{D0_theta}. The distribution is fitted here with the function $R_0(\theta)=61.7\:(1+3.43\:10^{-6}\theta^{3.22})$~m.

These distances are sensibly smaller than the experimental measurements, as for most current models, and this point needs to be clarified. However, the dependence over the zenith angle still is interesting.

\subsection{Global formula}
Different dependencies of the electric field have already been exposed:
\begin{itemize}
\item proportionality to the primary cosmic ray energy due to the number of particles
\item proportionality to the $-\mathbf{v}\times\mathbf{B}$ linked to the emission mechanism
\item a dependence on the zenith angle via $F(\theta)$, to interpret in terms of atmosphere and shower development
\item an exponential decrease with the distance of to the shower axis via $R_0(\theta)$
\end{itemize}
To generalize we can add a proportionality to the amplitude of the magnetic field\footnote{verified with ReAIRES but not presented here}. This gives the following formula for the electric field produced by an EAS around $10^{17}$~eV in a bandwidth of $\Delta\nu$~MHz:
\begin{eqnarray}
  \mathbf{E}&=&\frac{\Delta\nu}{60\textnormal{~MHz}}.\frac{|B|}{47\textnormal{~$\mu$T}}.\frac{Energy}{10^{17}\textnormal{~eV}}\nonumber\\
  &&{\times}\:F(\theta).\frac{-\mathbf{v}\times\mathbf{B}}{|v|.|B|}.\exp{\frac{-R}{R_0(\theta)}}
  \label{overall_formula}
\end{eqnarray}
where $F(\theta)=85.4\ (1+3.72\ 10^{-3})/(1+\exp \frac{\theta-6.32\ 10^{-3}}{2.17\ 10^{-3}})$~m.

%%%%%%%%%%%%%%%%%%%%%%%%%%%%%%%%%%%%%%%%%%%%%%%%%%%%%%%%%%%%%%%%%%%%%%%%%%
%%%%%%%%%%%%%%%%%%%%%%%%%%%%%%%%%%%%%%%%%%%%%%%%%%%%%%%%%%%%%%%%%%%%%%%%%%
\section{Discussion}
Synchrotron radio emission was implemented in AIRES to perform a full Monte Carlo simulation of the radio emission of cosmic rays EAS. We have detailed here an analysis and parameterization of the electric field simulated for different configurations. If some difference with experimental measurements still need to be understood (electric field too high, $R_0$ too small), important characteristics of the electric field can be extracted (as the $-\mathbf{v}\times\mathbf{B}$ polarization) and can guide experimental analysis, as for instance in~\cite{ICRC_Colas_codalma}.

%%%%%%%%%%%%%%%%%%%%%%%%%%%%%%%%%%%%%%%%%%%%%%%%%%%%%%%%%%%%%%%%%%%%%%%%%%
%%%%%%%%%%%%%%%%%%%%%%%%%%%%%%%%%%%%%%%%%%%%%%%%%%%%%%%%%%%%%%%%%%%%%%%%%%
\section*{Acknowledgement}
The authors would like to thank D. Kleckner, M.A. DuVernois and B. Cai for the radio version of AIRES.

%%%%%%%%%%%%%%%%%%%%%%%%%%%%%%%%%%%%%%%%%%%%%%%%%%%%%%%%%%%%%%%%%%%%%%%%%%%%%%%%%%%%%%%

\end{document}